\documentclass{esannV2}
\usepackage[dvips]{graphicx}
\usepackage[latin1]{inputenc}
\usepackage{amssymb,amsmath,array}
\usepackage{comment}
\usepackage{blindtext}
\usepackage{svg}
\usepackage{booktabs}
\usepackage{cite}
\usepackage{todonotes}
\usepackage{multirow}

\newcommand{\mol}{g}

\newcommand{\zb}{\mathbf{z}}

\newcommand{\Ab}{\mathbf{A}}
\newcommand{\Xb}{\mathbf{X}}
\newcommand{\Hb}{\mathbf{H}}

\newcommand{\inputspace}{\mathcal{X}}
\newcommand{\molsubset}{\mathcal{S}}

\newcommand{\energygradient}{\nabla_{\theta}}
\newcommand{\pprob}[1]{p_{\theta}(#1)}
\newcommand{\energy}[1]{E_{\theta}(#1)}

\newcommand{\denovo}{\textit{de novo }}

\newcommand{\modelname}{UEMO }
\begin{document}

\title{Towards Efficient Molecular Property Optimization with Graph Energy Based Models}
\author{Luca Miglior$^1$, Lorenzo Simone$^1$, Marco Podda$^1$ and Davide Bacciu$^1$
%
\thanks{This work is partially supported by the Next Generation EU programme under project FAIR (PE00000013), Spoke 1}
%
\vspace{.3cm}\\
%
1 - University of Pisa - Department of Computer Science \\
Largo Bruno Pontecorvo 3, Pisa, Italy
}

\maketitle

\begin{abstract}
Optimizing chemical properties is a challenging task due to the vastness and complexity of chemical space. Here, we present a  generative energy-based architecture for implicit chemical property optimization, designed to efficiently generate molecules that satisfy target properties without explicit conditional generation. We use Graph Energy Based Models and a training approach that does not require property labels. We validated our approach on well-established chemical benchmarks, showing superior results to state-of-the-art methods and demonstrating robustness and efficiency towards \denovo drug design.
\end{abstract}

\section{Introduction}
Generating molecules that satisfy specific chemical constraints is  crucial in modern \denovo drug design. Traditional generative approaches for property optimization generally involve training deep learning models on large datasets as general-purpose molecule generators. 
Prior works explored property optimization using Reinforcement Learning (RL) methods to guide molecular generation, such as GraphDF \cite{pmlr-v139-luo21a} and GraphAF \cite{Shi*2020GraphAF:}, or Bayesian Optimization (BO) on the latent space as proposed in JT-VAE \cite{pmlr-v80-jin18a}. MoFlow \cite{10.1145/3394486.3403104} uses Classifier-Based (CB) guidance to direct the generation towards specific objectives. GraphEBM \cite{liu2021graphebmmoleculargraphgeneration} tackles molecular generation with energy-based models (EBM), but the results of property optimization remain unclear, as it restricts to optimizing numerically quantifiable properties and exhibits high variance in generation, which affects consistency and reliability of generated molecules. Despite their advancements, the methodologies referenced above present several limitations: RL methods are known to be unstable and computationally demanding, while BO and CB methods rely on supervised learning and require large labeled datasets to work properly, which are rarely available in the chemical domain. 

Motivated by the need of overcoming existing limitations in optimization methods, especially in real-world scenarios where labeled data is scarce and computational efficiency is crucial, we present a novel Unspervised Energy-based Molecule Optimization (UEMO) capable of implicitly learning chemical properties and efficiently generating molecular graphs without the need of labeled data or explicit optimization strategies. Our method leverages the capabilities of Graph Neural Networks (GNNs) \cite{bacciuGentleIntroductionDeep2020} to process molecular graphs and utilizes Langevin dynamics \cite{duImplicitGenerationModeling2019} at sampling time to generate new molecules. We tested our approach on ZINC-250K \cite{irwin_zinc_2012}, evaluating its performance in optimizing QED and LogP chemical properties, and providing results on the constrained optimization of existing compounds.

\section{Methodology}
Whilst most models are trained as general-purpose generators, with property-specific optimization treated as a separate post hoc task, we exceed such limitations by proposing a novel method that implicitly embeds a strong bias toward the desired objective directly in the generative model, eliminating the need for supervision. Given a molecular input space \(\inputspace\), let \(\molsubset \subseteq \inputspace \) the subset of molecules that satisfy given property constraints. The goal is to train the generative model that maximizes the probability of generating samples coming from \(\molsubset\). Such a result can be pursued by resorting to domain-specific datasets guaranteed to reflect the targeted property (e.g., binding affinity to a specific property), or by employing off-the-shelf property predictors to filter molecules that satisfy desired constraints from existing datasets. This concept-based approach is versatile and applicable to any desired chemical objective. In contrast to baseline methods -- 
 GraphEBM \cite{liu2021graphebmmoleculargraphgeneration} in particular, whose optimization strategy is limited by numerical property estimates to scale the energy term -- we provide a generalizable setting that allows for the optimization of arbitrary properties without requiring explicit values or labeled data. This is particularly advantageous when dealing with complex or non-numerical targets, such as binding affinity or toxicity. 

\begin{figure}[t!]
    \centering
    \includegraphics[width=0.9\textwidth]{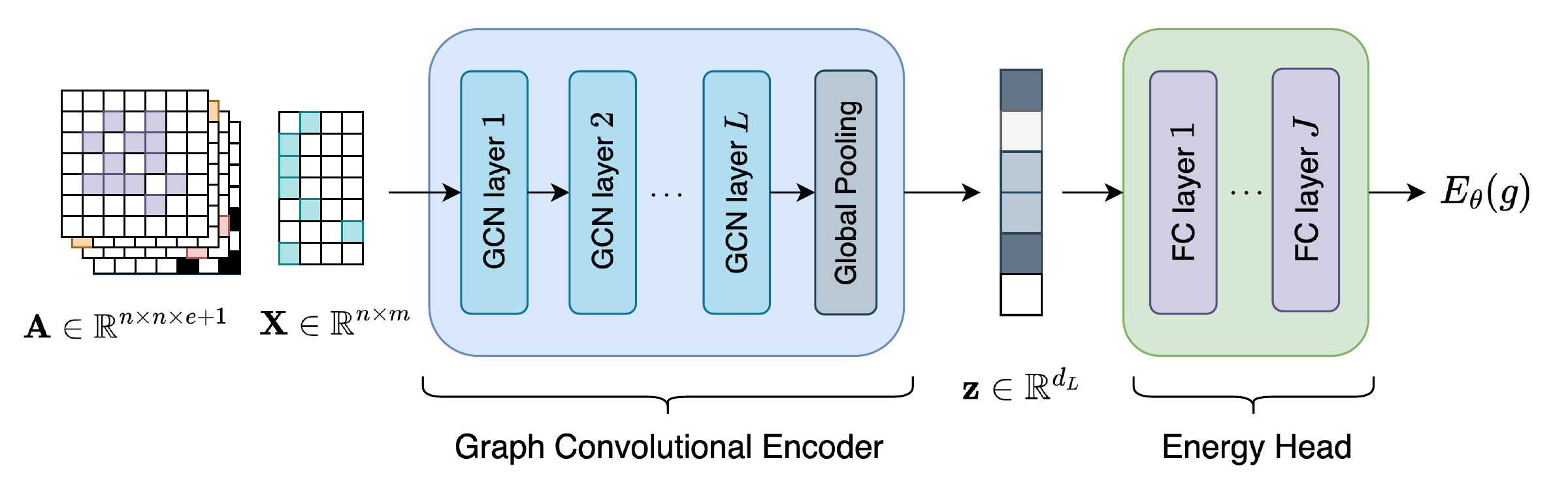} 
    \caption{\modelname architecture. In the example, $n=7, m=4,$ and $e=3$.}
    \label{fig:energy_model} 
\end{figure}

Let \(\mol = (\Ab, \Xb)\) be a molecular graph, consisting of \(n\) atoms and \(m\) atom types, where \(\Ab \in \mathbb{R}^{n \times n \times (e+1)}\) is the adjacency tensor describing chemical bonds\footnote{$e$ is the number of bond types, with an additional type to indicate the absence of a bond.}, and \(\Xb \in \mathbb{R}^{n \times m}\) is the one-hot-encoded atom representation. The energy function is defined as a function \( E_\theta(\mol) : \mathcal{X} \rightarrow \mathbb{R} \) with learnable parameters \(\theta\) that assigns a scalar value to each molecule in the input space.
The probability distribution over the input space is given by:
\[
p_\theta(\mol) = \frac{1}{Z} \exp(-E_\theta(\mol)),
\]
where \( Z = \int_{\mathcal{X}} \exp(-E_\theta(\mol)) \, d\mol \) is the intractable marginalization term. In this work, we model the energy function \( E_\theta \) using a Graph Convolutional Network (GCN) \cite{kipfSemiSupervisedClassificationGraph2017}. The GCN consists of \(L\) message-passing layers, 
after which a graph-level embedding \(\mathbf{z} \in \mathbb{R}^{d_L}\) is computed by applying a mean-pooling operation on the $L$-th node representations as follows:  \( \mathbf{z} = \text{MeanPooling}(\Hb^{(L)})\). 
Finally, the energy value is obtained by passing \(\zb\) through $J$ Fully Connected (FC) layers:
\[
\energy{\mol} = \text{EnergyHead}(\zb) = FC_J(FC_{J-1}(\ldots FC_1(\zb)))
\]
The schematic representation of the proposed architecture is shown in Figure \ref{fig:energy_model}.
Given the energy \(\energy{\mol}\in \mathbb{R}\), generating samples from the associated probability distribution is not trivial due to the intractability of the marginalization term. In this work, we use Langevin dynamics which directly uses the gradient of the energy function to generate samples:
\[
\mol_{i+1} = \mol_{i} - \frac{\epsilon}{2} \nabla_{\mol} E_{\theta}(\mol_{i}) + \omega_i, \, \omega_i \sim \mathcal{N}(0, \epsilon),
\]
where \(\epsilon \) is the step size. For \(i \rightarrow \infty \) and \(\epsilon \rightarrow 0\), this process is guaranteed to generate samples from the true distribution \(p_\theta\). \modelname is trained by modeling the energy of the space of the molecules \(\mathcal{X}\) through the distribution implicitly defined by \(E_\theta\). This is achieved by minimizing the negative log likelihood  \(\mathcal{L}(\theta) = -\mathbb{E}_{\mol \sim \molsubset}[\log \pprob{\mol}]\), under the empirical distribution of the dataset \(\molsubset\). This objective is known to have the formulation:
\[
\energygradient \mathcal{L}(\theta) = \mathbb{E}_{\mol^{+} \sim \molsubset} \left[ \energygradient \energy{\mol^{+}} \right] -\mathbb{E}_{\mol^{-} \sim p_{\theta}} \left[ \energygradient \energy{\mol^{-}} \right],
\label{eq:ebm_gradient}
\]
here we rely on Langevin dynamics to generate negative samples \( g^- \sim p_\theta\). Initial sampling hyperparameters for Langevin dynamics were chosen according to \cite[Appendix A.11]{duImplicitGenerationModeling2019}. As initial configuration for the chain \(\mol_1, \ldots \mol_i\ \), we leverage Persistent Contrastive Divergence (PCD) \cite{tielemanUsingFastWeights2009}, by implementing a limited-size replay buffer \(\mathcal{B}\) to store samples generated in previous iterations: in our implementation 95\% of the initializations are drawn from \(\mathcal{B}\) and 5\% from standard Gaussian noise. New molecules are sampled from the trained model by initializing the Langevin dynamics with molecular graphs \(g\) drawn from Gaussian noise, such that \(\Ab_0, \Xb_0 \sim \mathcal{N}(0,1)\). At each step of the sampling chain, the generated adjacency tensor is symmetrized to ensure consistency and used as initialization for the next step, following \(\Ab_{i+1} = \frac{1}{2}(\Ab_i + \Ab^T_i)\). Additionally, we compute the target property of the intermediate molecule at every step of the chain to maintain stability and prevent the local representation from collapsing. 

\section{Experiments}

We evaluated \modelname in two different property optimization tasks: Quantitative Drug Likeliness (QED) and Octanol-Water Partition Coefficient (LogP). To perform these experiments, we used the publicly available dataset ZINC-250k, which consists of 250,000 chemical compounds annotated with the above-mentioned properties. To facilitate implicit optimization, we divided the dataset in two partitions, each one reflecting a desired property constraint. Specifically, our partition matched high QED score (\(\geq\) 0.75) and high LogP coefficient (\(\geq\) 3) constraints. The partitioning was chosen to reflect real-world scenarios, where datasets targeting specific molecular properties are often small. The final partitions consisted of 54k and 34k molecules for QED and LogP, respectively. In addition to property evaluation, we performed benchmarks on molecular graph generation using three widely adopted metrics: validity, novelty, and uniqueness. These metrics are computed on 5,000 randomly generated molecules. We compared our methodology against five strong baselines \cite{10.1145/3394486.3403104, pmlr-v80-jin18a, Shi*2020GraphAF:, pmlr-v139-luo21a, liu2021graphebmmoleculargraphgeneration} by running their original source code and generating graphs with their property optimization strategies to ensure a fair comparison. JT-VAE results are from \cite{pmlr-v139-luo21a}. Firstly, we report results for the QED optimization task. We trained \modelname on the high-QED partition of the dataset and we randomly sampled a batch of molecules. Quantitative results are presented in Table \ref{tab:vertical_metrics_benchmarks}.
\begin{table}[ht]
    \centering
    \scriptsize
    \resizebox{0.89\textwidth}{!}{
    \begin{tabular}{lcccccc}
    \toprule
    \textbf{Model} & \textbf{Validity} & \textbf{Uniqueness} & \textbf{Novelty} & \textbf{Avg. QED} $\uparrow$ & \textbf{Top 1} $\uparrow$ \\
    \midrule
    GraphDF  & $100\%$ & $100\%$ & $100\%$ & $0.53 \pm 0.28$ & \textbf{0.948} \\
    GraphAF  & $100\%$ & $100\%$ & $100\%$ & $0.52 \pm 0.17$ & \textbf{0.948} \\
    GraphEBM & $100\%$ & $83.1\%\pm 2.7\%$  & $100\%$ & $0.44 \pm 0.11$ & 0.781 \\
    MoFlow   & $100\%$ & $96.8\% \pm 1.6\%$  & $100\%$ & $0.59 \pm 0.23$ & \textbf{0.948} \\
    JT-VAE   & $100\%$ & $100\%$ & $100\%$ & n.a. & 0.925 \\
    \midrule
    \textbf{\modelname} & \textbf{100\%} & \textbf{100\%} & \textbf{100\%} & \textbf{0.79 $\pm$ 0.12} & \textbf{0.948} \\
    \midrule
     &  &  &  & \textbf{Avg. LogP} $\uparrow$ & \textbf{Top 1} $\uparrow$ \\
    \midrule
    GraphDF  & $100\%$ & $100\%$ & $100\%$ & $9.13 \pm 1.00$ & 14.12 \\
    GraphAF  & $100\%$ & $100\%$ & $100\%$ & $8.15 \pm 3.90$ & \textbf{14.82} \\
    GraphEBM & $100\%$ & $82.6\%\pm 2.8\%$  & $100\%$ & $-1.50 \pm 2.34 $ & 3.29 \\
    MoFlow   & $100\%$ & $95.4\% \pm 1.6\%$  & $100\%$ & $-2.30 \pm 5.37 $  & 3.93 \\
    JT-VAE   & $100\%$ & $100\%$ & $100\%$ & n.a. & n.a. \\
    \midrule
    \textbf{\modelname} & \textbf{100\%} & \textbf{100\%} & \textbf{100\%} & \textbf{9.30 $\pm$ 2.13} & 12.66 \\
    \bottomrule
    \end{tabular}
    }
    \caption{\centering Metrics for molecules generated in both optimization tasks. }
    \label{tab:vertical_metrics_benchmarks}
\end{table}

\begin{figure}[t!]
    \centering
    \includegraphics[width=0.9\textwidth]{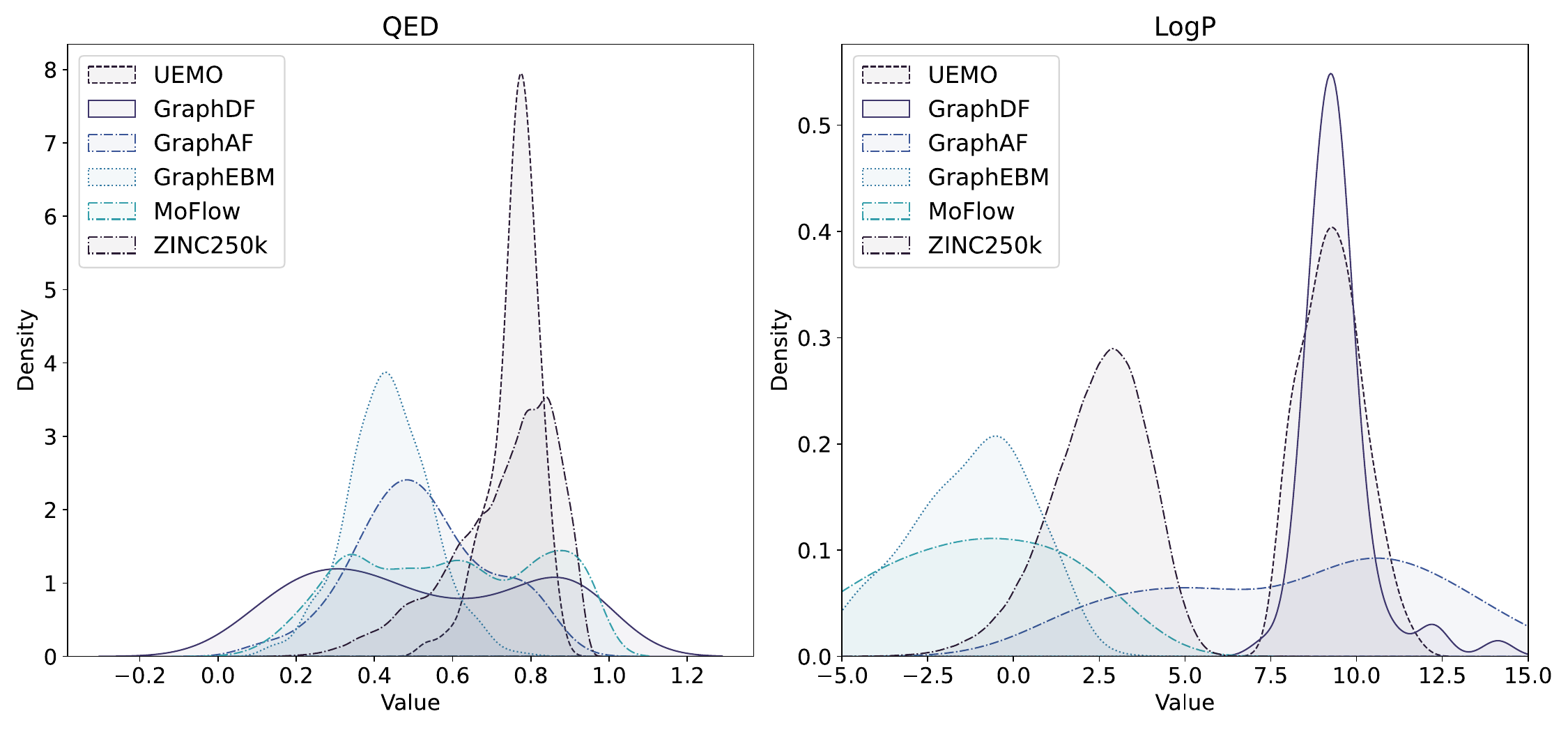}
    \caption{\centering QED (left) and LogP (right) distributions of the molecules generated by the different methods compared to the full, not partitioned, ZINC-250k dataset.}
    \label{fig:models_dist_comp}
\end{figure}

\noindent
Our model achieves perfect validity, uniqueness, and novelty ratios, matching the performance of all tested models except GraphEBM, which exhibits a substantially lower uniqueness ratio. Furthermore, our model significantly outperforms all other baselines with an average QED score of \(\mathbf{0.79\pm0.12}\), demonstrating that our implicit approach is more effective at sampling various and complex molecular configurations from the learned energy landscape. Additionally, it shows lower variance compared to the baselines, highlighting the robustness of our method in directing the generative process towards optimizing the desired chemical property. As illustrated in the right pane of Figure \ref{fig:models_dist_comp}, the Kernel Density Estimation (KDE) plot for the QED property of the generated molecules aligns closely with that of the training partition. This alignment confirms our model's ability to sample challenging molecules from a small, specific region of chemical space. The plot further emphasizes the stability of the generative process, validating the reliability of our implicit optimization strategy.
\sloppy{We then report LogP optimization results in Table \ref{tab:vertical_metrics_benchmarks} (bottom). Our experiments demonstrate that our model achieves performance comparable to state-of-the-art autoregressive strategies \cite{pmlr-v139-luo21a, Shi*2020GraphAF:}, generating molecules with an average LogP of \(\mathbf{9.30 \pm 2.13}\). Moreover, \modelname outperforms one-shot generative methods \cite{pmlr-v80-jin18a, 10.1145/3394486.3403104, liu2021graphebmmoleculargraphgeneration}, which either failed the task or produced molecules that did not satisfy the required constraints.}
Although maximizing LogP alone does not have direct practical applications and is considered a relatively straightforward task (since high LogP values can be trivially achieved by adding long chains of carbon atoms), 
success in this task highlights the flexibility of our proposed methodology across diverse and distinct generative tasks, effectively capturing chemical information from small and domain-specific datasets. This result can be visualized in Figure \ref{fig:models_dist_comp}, where the majority of  molecules generated by \modelname exhibit high LogP values with low variance, emphasizing the robustness of our approach. 
\begin{figure}[b]
    \centering
    \includegraphics[width=\linewidth]{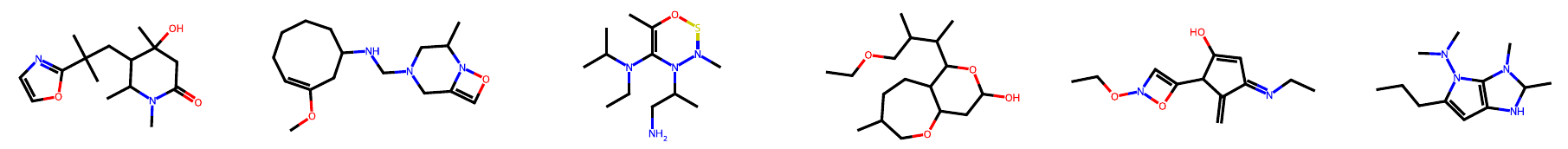}
    \caption{Curated samples of molecules generated in the QED optimization task.}
    \label{fig:cureted_samples}
\end{figure}
A visual representation of UEMO's generated molecules is presented in Figure \ref{fig:cureted_samples} for the QED optimization task.

While achieving state-of-the-art performance in property optimization tasks such as QED and LogP, \modelname also stands out for its remarkable sampling efficiency. Unlike baseline strategies, which require significant computational resources, our model demonstrates a highly efficient generative process. Notably, \modelname is able to generate molecules at an average rate of \(\mathbf{14.6 \pm 1.2 s/100\text{\textbf{mol}}}\). We provide a comprehensive comparison of sampling times in Table \ref{tab:sampling_times}. This substantial efficiency arises from two key factors. First, our implicit optimization strategy eliminates the need for expensive sampling strategies, which are employed by other methods. Second, \modelname  architecture is exceptionally lightweight, consisting in 0.25M trainable parameters, which is orders of magnitude smaller than many state-of-the-art models. This efficiency not only reduces computational overhead, but also makes \modelname particularly well suited for applications that require a rapid and focused exploration of the chemical space.

\begin{table}[t!]
    \centering
    \scriptsize
    \resizebox{0.9\textwidth}{!}{
    \begin{tabular}{cccccc}
    \toprule
    \textbf{\modelname} & \textbf{GraphDF} & \textbf{GraphAF} & \textbf{GraphEBM} & \textbf{MoFlow} & \textbf{JT-VAE} \\
    \midrule
    \textbf{14.6} $\mathbf{\pm 1.2}$ & $174 \pm 6.1 $  & $41.5\pm 5.3$ & $212.1 \pm 4.1$& $366 \pm 3.7 $  & n.a.\\
    \bottomrule
    \end{tabular}
    }
    \caption{\centering Time taken (in seconds) to generate 100 molecules.}
    \label{tab:sampling_times}
\end{table}

\section{Conclusions}
In this paper we presented UEMO, a novel EBM for molecular generation and chemical property optimization. Our implicit optimization approach demonstrated strong performance, outperforming existing models. Moreover, our lightweight model and innovative approach showed outstanding sampling efficiency, significantly reducing graph generation times. 
In the future, we plan to extend this work to larger and drug-specific datasets targeting diverse properties to generate highly specific graphs. Additionally, we plan to develop training methodologies that incorporate chemical knowledge into the energy landscape, ensuring a \textit{chemically informed} generative process that adheres to chemical rules.

\begin{footnotesize}

\bibliographystyle{abbrv}
\bibliography{bibliography}
\end{footnotesize}


\end{document}